\documentclass[a4paper,11pt]{article}

\usepackage[dvipsnames]{xcolor}
\usepackage{jheppub}
\usepackage{graphicx}
\usepackage{amsmath,amssymb,amsthm,amsfonts}
\usepackage{booktabs}
\usepackage{slashed}
\usepackage{braket}
\usepackage[export]{adjustbox}
\usepackage{enumitem}
\usepackage{placeins}
\usepackage[normalem]{ulem}
\usepackage{multirow}
\usepackage{hyperref}
\usepackage{mathrsfs}
\usepackage{dsfont}

\usepackage{verbatim}
\usepackage[T1]{fontenc}
\usepackage[utf8]{inputenc}
\usepackage{array}
\usepackage{wrapfig}
\usepackage{tabu}
\usepackage{longtable}
\usepackage{float}
\usepackage{caption}
\usepackage{subcaption}
\usepackage{tikz,tikz-3dplot}
\usetikzlibrary{calc,arrows.meta,shapes.geometric,decorations.markings,decorations.pathmorphing}
\usepackage{soul}
\usepackage[usestackEOL]{stackengine}
\usepackage{cancel}
\usepackage{tabularx}
\usepackage{makecell}

\allowdisplaybreaks

\definecolor{mycolor}{rgb}{0.710, 0.325, 0.165}

\definecolor{mycolor}{HTML}{2563EB}

\hypersetup{
    colorlinks=true,
    linkcolor=mycolor,
    citecolor=mycolor,
    filecolor=mycolor,
    urlcolor=mycolor
    }

\def\beq{\begin{equation}}
\def\eeq{\end{equation}}
\def\bea{\begin{eqnarray}}
\def\eea{\end{eqnarray}}
\def\df{\text{d}}

\renewcommand{\imath}{\mathrm{i}}
\newcommand{\alphas}{\alpha_{\mathrm{s}}}

\providecommand{\preprint}[1]{}

\preprint{SLAC-PUB-260710}

\title{Three-Loop {\color{red}Q}{\color{green}C}{\color{blue}D} Corrections to Scattering Amplitudes of a Higgs Boson and Three Partons}

\author{Xin Guan,}
\emailAdd{guanxin@slac.stanford.edu}

\author{Bernhard Mistlberger,}
\emailAdd{bernhard.mistlberger@gmail.com}

\author{Michael S. Ruf}
\emailAdd{mruf@slac.stanford.edu}

\affiliation{SLAC National Accelerator Laboratory, Stanford University, Stanford, CA 94039, USA}

\abstract{
We present helicity amplitudes at three loops in the heavy-top quark effective theory of QCD for the scattering of a Higgs boson and three light partons with full-color dependence.
To obtain these results, we use integration-by-parts identities implemented in the code Blade and evaluate the master integrals using canonical differential equations.
We obtain compact analytic expressions in terms of generalized polylogarithms suitable for numerical evaluation.
We observe that sub-leading color contributions at three loops are numerically as significant as leading-color terms.
Our amplitudes are key ingredients for precision LHC phenomenology, for example in production cross sections involving a Higgs boson and a hadronic jet.
}

\begin{document}
\maketitle
\vspace{0.5em}

\newpage
\section{Introduction}

Higgs boson phenomenology is one of the most exciting fields of physics of our time.
The Large Hadron Collider (LHC), with the discovery of the Higgs boson by the ATLAS~\cite{ATLAS:2012yve} and CMS~\cite{CMS:2012qbp} collaborations, opened the window to the domain of ultra-short-ranged interactions that allow us for the first time to probe this fundamental particle directly.
The results derived from these LHC measurements hold the promise to deepen our fundamental understanding of nature and unravel some of the most puzzling mysteries of physics.
Achieving these lofty goals, in particular with the advent of the High-Luminosity phase of the LHC~\cite{ZurbanoFernandez:2020cco,ATLAS:2025eii,Cepeda:2019klc}, has created a thirst for highly precise predictions for the outcome of LHC measurements.
In this article, we take a leap towards quenching this thirst and present the computation of three-loop QCD corrections to scattering amplitudes involving the Higgs boson and three partons.

We pursue a computation to ultimately advance our predictions for a Higgs boson coupling to QCD partons via an effective field theory (EFT) in which we consider the degrees of freedom of the top quark to be infinitely heavy~\cite{Wilczek1977,Shifman1978,Inami1983,Spiridonov:1988md}.
The currently highest-accuracy predictions for the total rate of producing Higgs bosons via the gluon-fusion production mechanism at the LHC (the overwhelmingly dominant mode to produce the Higgs) have been obtained using this EFT~\cite{Anastasiou:2015vya,Anastasiou:2016cez,Mistlberger:2018etf}.
These predictions were carried out at next-to-next-to-next-to-leading order (N$^3$LO) in QCD perturbation theory and form the current state of the art alongside their lower-order counterparts~\cite{Dawson:1990zj,Djouadi:1991tka,Anastasiou:2002yz,Harlander:2002wh,Ravindran:2003um} and crucial mass~\cite{Czakon:2020vql,Czakon:2021yub,Czakon:2023kqm,Czakon:2024ywb,Graudenz:1992pv,Spira:1995rr} and electroweak corrections~\cite{Actis:2008ts,Actis:2008ug,Aglietti:2004nj,Becchetti:2020wof}.
Despite these tremendous efforts, the current state of the art will not be enough for the precision goals of the HL-LHC, and already now theoretical uncertainties dominate our ability to extract additional information from experimental observations.

 The diverse range of observables measurable at the LHC has sparked a program that goes far beyond the total rate of observed Higgs bosons.
 The theoretical community, as a consequence, has produced tools and highly complex computations to better interpret the experimental outcome in the very same EFT.
 Predictions based on parton showers are available through NNLO~\cite{Hoche:2014dla,Monni:2019whf,Alioli:2023har}.
Differential predictions for distributions of the Higgs boson at N$^3$LO accuracy were presented for example in Refs.~\cite{Dulat:2018bfe,Chen:2021isd,Cieri:2018oms}.
Even more intricate final states, such as the production of a Higgs boson alongside a hadronic jet, are currently available at NNLO~\cite{Boughezal:2015dra,Boughezal:2015aha,Chen:2014gva,Caola:2015wna,Chen:2016zka}.
Importantly, the hadronic jet can provide recoil to the Higgs boson and boost its final states transverse to the LHC collision axis.
Consequently, precision on this observable is of high importance.

Our computation of the Higgs plus three parton amplitudes we present in this manuscript serves as a direct input to improve future predictions of the aforementioned observables.
Our amplitudes take the role of purely virtual corrections to Higgs boson production in gluon-fusion in association with a jet at N$^3$LO.
Furthermore, after inclusive integration over phase space, they are a crucial ingredient to the N$^4$LO gluon fusion production cross section.
First steps to achieve this result are already underway~\cite{Das:2020adl,Lee:2021uqq,Lee:2022nhh} and in particular contributions with one final-state parton interfering two- and one-loop amplitudes are already available~\cite{Mistlberger:2025ksa}.
Additionally, our amplitudes will play an important role in the prediction of the decay of the Higgs boson to jet final states.

Our computation in this article builds on a significant history of results.
The computation of the required Feynman integrals and scattering amplitudes at two loops~\cite{Gehrmann:2000zt,Gehrmann:2011aa,Gehrmann:2023etk,Gehrmann:2001ck,Duhr:2014nda} was at the beginning of exploring more intricate function classes~\cite{Gehrmann:2001jv} in high-energy phenomenology beyond so-called harmonic polylogarithms~\cite{Remiddi:1999ew}.
The amplitudes at two-loop order are known at higher powers in the dimensional regularization parameter $\epsilon$~\cite{Gehrmann:2023etk}.
The three-loop function space has been studied in detail in Ref.~\cite{Gehrmann:2024tds,DiVita:2014pza,Canko:2021xmn,Henn:2023vbd,Syrrakos:2023mor,Guan:2025awp}.
Three-loop corrections to the amplitude interfered with its tree-level counterpart, computed in the generalized leading-color approximation, were presented in Ref.~\cite{Chen:2025utl}.
In addition, scattering amplitudes relying on the same integrals to express them for the scattering of an off-shell vector boson and three partons were computed in Ref.~\cite{Gehrmann:2023jyv} in the generalized leading-color approximation.

In addition to the pursuit of the amplitudes considered here, considerable interest in their counterpart in maximally supersymmetric Yang-Mills ($\mathcal{N}=4$ sYM) theory has developed.
The form factor for the stress-tensor multiplet, represented by the operator $\operatorname{tr}(\phi^2)$, is a direct cousin of the Higgs-plus-three-gluon amplitudes.
A striking observation is that the leading transcendental part of this object in this theory and in QCD is identical~\cite{Brandhuber:2012vm} at two loops.
This correspondence was recently confirmed at the three-loop level in the leading-color limit in Ref.~\cite{Chen:2025utl}.
By virtue of the amplitude bootstrap, the three-point form factor was recently computed through astounding eight-loop order~\cite{Dixon:2020bbt,Dixon:2022rse}.
The three-loop result has been verified through an explicit analytic calculation in Ref.~\cite{Gehrmann:2024tds}.
The form factor was recently computed beyond the leading-color limit in Ref.~\cite{Guan:2025awp} by the authors.
Here, we compare the QCD result to this computation and find that the leading transcendental part of our amplitudes indeed corresponds to the result in $\mathcal{N}=4$ sYM theory.

In our computation we generated Feynman diagrams for the required scattering amplitudes through three-loop order and project them onto gauge-invariant parts of the scattering amplitude~\cite{Gehrmann:2013vga,Gehrmann:2011aa,Peraro:2020sfm,Goode:2024cfy,Goode:2024mci}.
We then use integration-by-parts (IBP) identities~\cite{Chetyrkin1981,Tkachov1981} to relate our loop integrals to a finite basis of so-called master integrals, in particular implemented in the program Blade~\cite{Guan:2024byi}.
We use results from the literature for these master integrals~\cite{Gehrmann:2024tds,DiVita:2014pza,Canko:2021xmn,Henn:2023vbd,Syrrakos:2023mor}, the computation of missing master integrals for the purposes of Ref.~\cite{Guan:2025awp}, and we compute additional missing master integrals using the method of differential equations~\cite{Henn:2013pwa,Gehrmann:1999as,Kotikov:1990kg,Kotikov:1991hm,Kotikov:1991pm} in combination with regularity conditions to fix their boundary conditions, see for example Refs.~\cite{Henn:2020lye,Dulat:2014mda,Henn:2013woa}.
We study the infrared structure of our amplitudes, validate universal infrared~\cite{Herzog:2023sgb} and collinear~\cite{Guan:2024hlf} limits and compare to the literature.
Our final results are scattering amplitudes valid in conventional dimensional regularization as well as the 't~Hooft-Veltman scheme for all processes involving a Higgs boson and three partons expressed in terms of linear combinations of generalized polylogarithms and algebraic coefficients.
Our results are openly available~\cite{ZenodoResults}.

The remainder of this paper is organized as follows.
In Sec.~\ref{sec:Setup}, we describe the kinematic setup and the decomposition of the amplitudes into gauge-invariant Lorentz and helicity structures.
In Sec.~\ref{sec:Computation}, we detail the computation, focusing on the integrand construction, the integration-by-parts reduction, and the evaluation of the
master integrals. In Sec.~\ref{sec:IR}, we discuss the infrared structure of the amplitudes, which we exploit both as a check and to define the finite remainders.
We present and analyze our results in Sec.~\ref{sec:Results}.
We describe the validation of our computation in Sec.~\ref{sec:Validation}.
Finally, we present our conclusions in Sec.~\ref{sec:Conclusions}.

\section{Set-Up \label{sec:Setup}}
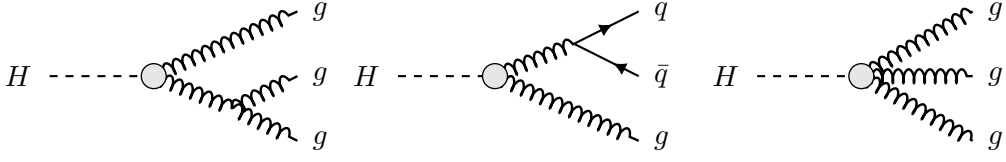
\begin{figure}[t]
    \centering
    \begin{tikzpicture}[scale=0.95]
        \tikzset{
            sline/.style={dashed,thick},
            fline/.style={thick,postaction={decorate},decoration={markings,mark=at position 0.60 with {\arrow{Latex[length=5pt,width=5pt]}}}},
            afline/.style={thick,postaction={decorate},decoration={markings,mark=at position 0.60 with {\arrowreversed{Latex[length=5pt,width=5pt]}}}},
            gline/.style={draw=black,line width=0.9pt,decorate,decoration={coil,aspect=0.35,segment length=1.4mm,amplitude=0.9mm}},
            hblob/.style={circle,fill=gray!20,draw=black,line width=0.5pt,minimum size=3.3mm,inner sep=0pt},
            hblobfront/.style={circle,fill=gray!20,draw=black,line width=0.5pt,minimum size=3.3mm,inner sep=0pt}
        }
        
        \coordinate (h1) at (0.35,0);
        \coordinate (v1) at (1.8,0);
        \coordinate (v1g) at (2.9,-0.45);
        \coordinate (g1u) at (3.8,0.9);
        \coordinate (g1m) at (3.8,0.0);
        \coordinate (g1d) at (3.8,-0.9);
        \draw[sline] (h1) -- (v1);
        \draw[gline] (v1) -- (g1u);
        \draw[gline] (v1) -- (v1g);
        \draw[gline] (v1g) -- (g1m);
        \draw[gline] (v1g) -- (g1d);
        \node[left=3pt] at (h1) {$H$};
        \node[right=2pt] at (g1u) {$g$};
        \node[right=2pt] at (g1m) {$g$};
        \node[right=2pt] at (g1d) {$g$};

        \coordinate (h2) at (5.2,0);
        \coordinate (v2) at (6.55,0);
        \coordinate (split2) at (7.65,0.45);
        \coordinate (q2) at (8.55,0.9);
        \coordinate (qb2) at (8.55,0.0);
        \coordinate (g2d) at (8.55,-0.9);
        \draw[sline] (h2) -- (v2);
        \draw[gline] (v2) -- (g2d);
        \draw[gline] (v2) -- (split2);
        \draw[fline] (split2) -- (q2);
        \draw[afline] (split2) -- (qb2);
        \node[left=3pt] at (h2) {$H$};
        \node[right=2pt] at (g2d) {$g$};
        \node[right=2pt] at (q2) {$q$};
        \node[right=2pt] at (qb2) {$\bar q$};

        \coordinate (h3) at (10.2,0);
        \coordinate (v3) at (11.65,0);
        \coordinate (g3u) at (13.2,0.9);
        \coordinate (g3m) at (13.2,0.0);
        \coordinate (g3d) at (13.2,-0.9);
        \draw[sline] (h3) -- (v3);
        \draw[gline] (v3) -- (g3u);
        \draw[gline] (v3) -- (g3m);
        \draw[gline] (v3) -- (g3d);
        \node[hblobfront] at (v1) {};
        \node[hblob] at (v2) {};
        \node[hblob] at (v3) {};
        \node[left=3pt] at (h3) {$H$};
        \node[right=2pt] at (g3u) {$g$};
        \node[right=2pt] at (g3m) {$g$};
        \node[right=2pt] at (g3d) {$g$};
    \end{tikzpicture}
    \caption{Leading-order diagrams for Higgs decay into three partons. Higgs-gluon interactions are indicated by a gray blob. The dashed line denotes a Higgs boson, the coiled line a gluon, and the solid line with arrows a fermion.}
    \label{fig:LODiagrams}
\end{figure}

\noindent
We consider processes involving a Higgs boson $H$ with momentum $q$ and three light partons, either gluons $g$ or quarks $q$, with momenta $p_1$, $p_2$ and $p_3$, respectively.
To be concrete, we are interested in the processes
\begin{align}
    H(q)+ q(p_1)+\bar q(p_2)+g(p_3)\,&\to 0,\\
    H(q)+ g(p_1)+g(p_2)+g(p_3)\,&\to 0.
\end{align}
For convenience of our computation we treat the three partons as incoming and obtain all relevant scattering configurations via crossing symmetry.
The tree-level Feynman diagrams corresponding to these processes are shown in Fig.~\ref{fig:LODiagrams}.
Specifically, we work in an approximation to the Standard Model where the degrees of freedom of the top quark are integrated out~\cite{Wilczek1977,Inami1983,Spiridonov:1988md,Shifman1978} and we consider all remaining quarks to be massless,
\beq
p_1^2=p_2^2=p_3^2=0,\qquad q^2=m_H^2\,.
\eeq
This theory is specified via an effective Lagrangian
\beq
\mathcal{L}_{\mathrm{eff}}
=
\mathcal{L}_{\mathrm{QCD},5}
-\frac{1}{4}C^0\,HG^a_{\mu\nu}G_a^{\mu\nu}\,
\label{eq:Lagrangian}
\eeq
where $\mathcal{L}_{\mathrm{QCD},5}$ denotes the five-flavor QCD Lagrangian and $C^0$ is the Wilson coefficient~\cite{Chetyrkin:1997un,Schroder:2005hy,Chetyrkin:2005ia,Kramer:1996iq,Kataev:1981gr} of the effective Higgs-gluon operator.
\noindent
We introduce the dimensionless Mandelstam invariants
\beq
s=\frac{(p_1+p_2)^2}{q^2},\qquad
t=\frac{(p_2+p_3)^2}{q^2},\qquad
u=\frac{(p_1+p_3)^2}{q^2}=1-s-t.
\eeq
The gauge group of QCD is $\mathrm{SU}(N_c)$ and we choose to represent the color dependence of our amplitudes in terms of Casimir invariants,  generators of representation R $T_R^a$
and adjoint structure constants $f^{abc}$. Specifically, we define
\beq
[T_R^a,T_R^b]=\mathrm{i}f^{abc}T_R^c,\qquad
\mathrm{Tr}(T_R^aT_R^b)=T_R\,\delta^{ab}.
\eeq
The quadratic Casimirs are defined by
\beq
(T_R^{a}T_R^a)_{ij}=C_R \delta_{ij}, \hspace{1cm} T^a_{A, ij}=-\imath f^{aij}.
\eeq
Here, $R\in \{A, F\}$ indicates the adjoint or fundamental representation.
Furthermore, we define the fully symmetric tensors
\beq
d_R^{(n)\, a_1a_2\dots a_n}=\frac{1}{n!} \sum\limits_{\text{perm. } \sigma} \text{tr}\left[T_R^{a_\sigma(1)}T_R^{a_{\sigma(2)}} \dots T_R^{a_{\sigma(n)}} \right].
\eeq
With this, we define the cubic and quartic Casimir invariants
\beq
C^{(3)}_{RR^\prime}=d_R^{(3)\, a_1a_2a_3}d_{R^\prime}^{(3)\, a_1a_2a_3}, \hspace{1cm}
C^{(4)}_{RR^\prime}=d_R^{(4)\, a_1a_2a_3a_4}d_{R^\prime}^{(4)\, a_1a_2a_3a_4}.
\eeq
Specifically, for $\operatorname{SU}(N_c)$ the quadratic Casimirs are given by
\beq
\begin{aligned}
C_A &= N_c, &
C_F &= \frac{N_c^2-1}{2N_c}.
\end{aligned}
\eeq
The cubic Casimirs are given by
\beq
\begin{aligned}
C^{(3)}_{FF} = \frac{(N_c^2-1)(N_c^2-4)}{16 N_c}\,,\quad C^{(3)}_{AF} = 0\,,\quad 
C^{(3)}_{AA} = 0.
\end{aligned}
\eeq
The quartic Casimirs are
\beq
\begin{aligned}
C^{(4)}_{AA} &= \frac{(N_c^2-1)N_c^2(N_c^2+36)}{24}\,,\\
C^{(4)}_{AF} &= \frac{(N_c^2-1)N_c(N_c^2+6)}{48}\,,\\
C^{(4)}_{FF} &= \frac{(N_c^2-1)(N_c^4-6N_c^2+18)}{96 N_c^2}\,.
\end{aligned}
\eeq
The Dynkin indices are given by 
\beq
\qquad T_F=\frac{1}{2},\qquad T_A=C_A.
\eeq
We express our scattering amplitudes in terms of a linear combination of  scalar partial amplitudes.
\bea
\label{eq:ampdef}
\mathcal{A}_{H ggg}(s,t,u)&=& \imath g_{\mathrm{s}}C^0  \epsilon^a_\mu(p_1)\epsilon^b_\nu(p_2)\epsilon^c_\rho(p_3) f^{abc} \sum\limits_{i} T_i^{\mu\nu\rho}A^{H  ggg}_i(s,t,u).\\
\mathcal{A}_{H q\bar q g}(s,t,u)&=& \imath g_{\mathrm{s}} C^0  \epsilon^a_\mu(p_3) \bar v_ {c_j  s_j} (p_2) \left[ T_{F\, c_j c_k}^{a} \sum\limits_{i} T_{i\, s_j s_k}^{\mu}A^{H q\bar q g}_i(s,t,u) \right] u_{c_k  s_k} (p_1).\nonumber
\eea
Above, the $T_i$ represent Lorentz tensor structures, $g_{\mathrm{s}}$ is the strong coupling constant, $\epsilon^\mu(p)$ is a gluon polarization vector and $u$ and $\bar v$ are the spinors of an incoming quark and anti-quark.

We discuss the decomposition of the amplitudes into gauge-invariant Lorentz tensor structures as well as in terms of specific helicity states in more detail in dedicated subsections below.
The scalar partial amplitudes can then be expanded in the strong coupling constant as follows.
\beq
\label{eq:pertexp}
A_i^ {H X}
=\sum_{n=0}^{\infty}a_{\mathrm{s}}^n\,{A}^{(n)\, H X}_i,\qquad
a_{\mathrm{s}}=\frac{\alphas}{\pi} e^{-\epsilon\gamma_{\mathrm{E}}}(4\pi)^{\epsilon},\qquad \alphas=\frac{g_{\mathrm{s}}^2}{4\pi}.
\eeq
Above, $X$ is either $ggg$ or $q\overline{q}g$ and
we have implicitly introduced that we use the $\overline{\mathrm{MS}}$-scheme of dimensional regularization in $d=4-2\epsilon$ dimensions throughout our computation.
Results for scattering channels are obtained by crossing from the decay kinematics and analytic continuation of the functional representation.
The explicit result of this article is the complete computation of the third-order partial scalar scattering amplitude $A_i^{(3),\, H+ X}$ and we elaborate on its computation and structure in subsequent sections.

\subsection{Lorentz Tensor Decomposition of the $Hggg$ Amplitude}

To specify the physical part of the scattering amplitude for a Higgs boson and three gluons, we choose the following gauge conditions:
\bea
\label{eq:ggggauge}
\epsilon^{\mu}_i \, p_{i,\,\mu}&=&0\,\nonumber\\
\epsilon^{\mu}_i \, p_{i+1,\,\mu}&=&0,\hspace{1cm}p_{4}=p_1,
\eea
where we omit the color indices and use $\epsilon^{\mu}_i$ as shorthand for $\epsilon^{\mu}(p_i)$.

To find Lorentz invariant tensor structures $T_i^{\mu\nu\rho}$ that span the physical space of our amplitude, we start with a generic ansatz of our scattering amplitude and then project onto the physical subspace making use of the gauge we defined, see Refs.~\cite{Gehrmann:2013vga,Gehrmann:2011aa,Peraro:2020sfm,Goode:2024cfy,Goode:2024mci} for recent literature on tensor decomposition of scattering amplitudes.
For a compact definition of our tensor structures, we introduce the following Lorentz tensors
\bea
P_{ijk}^{\mu}&=& \frac{p_j^\mu}{s_{ij}}-\frac{p_k^\mu}{s_{ik}}+\frac{s_{jk}}{s_{ij} s_{ik}}p_i^\mu .\nonumber\\
G_{ijk}^{\mu\nu}&=&g^{\mu\nu} -2 \frac{ p_j^\mu p_i^\nu}{s_{ij}} -2  \frac{p_k^\mu p_j^\nu}{s_{jk}} +2 s_{ik} \frac{p_j^\mu p_j^\nu}{s_{ij}s_{jk}}.
\eea
We find that we can express the entire physical scattering amplitude in terms of only four gauge-invariant tensor structures.
\bea
\label{eq:tensdefggg}
T_1^{\mu\nu\rho}&=&-q^2\frac{ (s+t) (t+u)}{t}G^{\nu\rho}_{231} P_{123}^\mu.\\
T_2^{\mu\nu\rho}&=& -q^2\frac{ (s+u) (t+u)}{u} G^{\mu\rho}_{213} P_{231}^\nu.\nonumber\\
T_3^{\mu\nu\rho}&=& -q^2\frac{ (s+t) (s+u)}{s} G^{\mu\nu}_{123}P_{312}^\rho.\nonumber\\
T_4^{\mu\nu\rho}&=&2 (q^2)^2 (st+su+tu) P_{123}^\mu P_{231}^\nu P_{312}^\rho.\nonumber
\eea
Explicitly enforcing the gauge conditions of Eq.~\eqref{eq:ggggauge} we find
\bea
\epsilon_\mu(p_1)\epsilon_\nu(p_2)\epsilon_\rho(p_3)T_1^{\mu\nu\rho}&=&\frac{ (s+t) (t+u)}{t u}(\epsilon_2\cdot \epsilon_3 )(p_3\cdot \epsilon_1),\\
\epsilon_\mu(p_1)\epsilon_\nu(p_2)\epsilon_\rho(p_3)T_2^{\mu\nu\rho}&=& \frac{ (s+u) (t+u)}{s u} (\epsilon_1\cdot \epsilon_3 )(p_1\cdot \epsilon_2),\nonumber\\
\epsilon_\mu(p_1)\epsilon_\nu(p_2)\epsilon_\rho(p_3)T_3^{\mu\nu\rho}&=& \frac{ (s+t) (s+u)}{s t} (\epsilon_1\cdot \epsilon_2 )(p_2\cdot \epsilon_3),\nonumber\\
\epsilon_\mu(p_1)\epsilon_\nu(p_2)\epsilon_\rho(p_3)T_4^{\mu\nu\rho}&=&  -\frac{2}{q^2} \frac{(st+su+tu)}{stu} (p_3 \cdot \epsilon_1 )(p_1\cdot \epsilon_2)(p_2\cdot \epsilon_3).\nonumber
\eea
We find that with the above normalization our tree-level partial scalar scattering amplitudes are normalized to one,
\beq
A_i^{(0)\,Hggg}=1.
\eeq

The tensor decomposition we introduced above holds in conventional dimensional regularization for arbitrary spacetime dimension.
Next, we introduce a parametrization of our scattering amplitude in terms of four-dimensional helicity prefactors in the 't~Hooft-Veltman scheme.
To this end we first define the positive and negative helicity components of the incoming gluon polarization vector.~\footnote{See for example Refs.~\cite{Dixon:2013uaa,Elvang:2015rqa} for an introduction to spinor-helicity methods.}
\beq
\label{eq:gluonpol}
\epsilon_{i,-}^\mu=-\frac{[p_{i+1}|\gamma^\mu | p_i \rangle }{\sqrt{2}[p_{i+1}   p_i]},\hspace{1cm}
\epsilon_{i,+}^\mu=\frac{\langle p_{i+1}|\gamma^\mu | p_i ] }{\sqrt{2} \langle p_{i+1}   p\rangle}.
\eeq

Above, the vector $\eta$ is the gauge vector associated with the gluon polarization vector $\epsilon$.
With this, the projection of an explicit helicity state of our amplitude is given in terms of partial helicity amplitudes.
\beq
 \epsilon_{\lambda_1,\, \mu}(p_1)\epsilon_{\lambda_2,\, \nu}(p_2)\epsilon_{\lambda_3,\,\rho}(p_3)  \sum\limits_{i} T_i^{\mu\nu\rho}A^{H ggg}_i(s,t,u)=\frac{1}{\sqrt{2}\, q^2 s t u}H^{Hggg}_{ \lambda_1\lambda_2\lambda_3}A^{Hggg}_{ \lambda_1\lambda_2\lambda_3}(s,t,u).
\eeq
Note that in the above equation we suppressed color dependence for convenience.
The helicity prefactors $H^{H ggg}_{ \lambda_1\lambda_2\lambda_3}$ are defined by
\bea
H_{---}&=&-H_{+++}^*=-\langle 12\rangle\langle 13\rangle\langle 23\rangle\,,\\
H_{--+}&=&-H_{++-}^*=s^2 \frac{ \langle 12\rangle^2}{[21]}[ 31][ 32]\,,\\
H_{-+-}&=&-H_{+-+}^*=u^2 \frac{ \langle 13\rangle^2}{[31]}[ 21][ 32]\,,\\
H_{+--}&=&-H_{-++}^*=t^2 \frac{ \langle 23\rangle^2}{[32]}[ 31][ 21]\,.
\eea
The partial helicity amplitudes $A^{H  ggg}_{ \lambda_1\lambda_2\lambda_3}(s,t,u)$ satisfy a perturbative expansion defined as in Eq.~\eqref{eq:pertexp}.
The tree-level amplitudes $A^{(0),\,H  ggg}_{ \lambda_1\lambda_2\lambda_3}(s,t,u)$ are normalized to 1,
\beq
A^{(0),\, H  ggg}_{ \lambda_1\lambda_2\lambda_3}(s,t,u)=1\,.
\eeq
Notably, only four of the eight helicity amplitudes are independent. In particular, we find
\bea
A^{H ggg}_{+++}&=&A^{H ggg}_{---}\,,\hspace{1cm}
A^{H ggg}_{++-}=A^{H ggg}_{--+}\,,\\
A^{H ggg}_{+-+}&=&A^{H ggg}_{-+-}\,,\hspace{1cm}
A^{H ggg}_{-++}=A^{H ggg}_{+--}\,.\nonumber
\eea
The partial helicity amplitudes form a particularly convenient basis for form factors as they enjoy a large range of symmetries.
The all-plus (all-minus) amplitude is completely symmetric under the exchange of $s$, $t$ and $u$.
The single-minus (single-plus) amplitudes are related to each other in the following way.
\bea
A^{H ggg}_{++-}(s,t,u)=A^{H ggg}_{-++}(u,s,t)=A^{H ggg}_{+-+}(t,u,s)\,.
\eea
Consequently, only two of our four form factors are independent functions.
We choose to present our final results in terms of the partial helicity amplitudes. For completeness, we give here the invertible relation to the form factors in the conventional dimensional regularization representation here.
\beq
\label{eq:hggghelamps}
\left(\begin{array}{c}
A^{H ggg}_{+++} \\ A^{H ggg}_{++-} \\ A^{H ggg}_{+-+} \\ A^{H ggg}_{-++}
\end{array}
\right)=
\left(
\begin{array}{cccc}
 (s+t) (t+u) & (s+u) (t+u) & (s+t) (s+u) & -s t-s u-t u \\
 0 & 0 & \frac{(s+t) (s+u)}{s^2} & -\frac{s t+s u+t u}{s^2} \\
 0 & \frac{(s+u) (t+u)}{u^2} & 0 & -\frac{s t+s u+t u}{u^2} \\
 \frac{(s+t) (t+u)}{t^2} & 0 & 0 & -\frac{s t+s u+t u}{t^2} \\
\end{array}
\right)
\cdot
\left(\begin{array}{c}
A^{H ggg}_1 \\ A^{H ggg}_2 \\ A^{H ggg}_3 \\ A^{H ggg}_4
\end{array}
\right).
\eeq

\subsection{Lorentz Tensor Decomposition of the $H q\bar q g$ Amplitude}

To specify the physical part of the scattering amplitude for a Higgs boson, a quark-antiquark pair and a gluon, we choose the following gauge conditions.
\bea
\label{eq:qqbarggauge}
\epsilon^{\mu}_3p_{3,\,\mu}&=&\epsilon^{\mu}_3p_{1,\,\mu}=0.
\eea
As in the gluon case, we start with an ansatz for a general Lorentz tensor for our scattering amplitude and then determine the physical tensor structures compatible with our gauge condition.
We find that the scattering amplitude can be represented in terms of only two tensor structures
\bea
\label{eq:tensdefqqbarg}
T_{1}^{\mu}&=&-\frac{1}{q^2 u } \slashed{p}_{3} \left[\frac{t}{s}p_1^\mu-\frac{u}{s} p_2^\mu +p_3^\mu \right],\\
T_{2}^{\mu}&=&-\frac{t+u}{2 s} \left[ \gamma^\mu_{}-\frac{2 }{q^2 u}\slashed{p}_{3} p_1^\mu \right].\nonumber
\eea

We find that with the above normalization our tree-level partial scalar scattering amplitudes are normalized to one,
\beq
A_i^{(0)\,H q\bar q g}=1.
\eeq

Using the definition of the four-dimensional helicity components of the gluon polarization of Eq.~\eqref{eq:gluonpol} and the assignment of the positive and negative helicity components of the spinors
\beq
u_+(p)=|p],\hspace{1cm}
u_-(p)=|p\rangle,\hspace{1cm}
\bar v_-(p)=\langle p| ,\hspace{1cm}
\bar  v_+(p)=[ p|,
\eeq
we can express the partial scalar amplitudes in terms of a helicity prefactor and partial helicity amplitudes.
\beq
 \epsilon_{\lambda_3\, \mu}(p_3) \bar v_ {\lambda_2\, s_j} (p_2)  \sum\limits_{i} T_{i\, s_j s_k}^{\mu}A^{H q\bar q g}_i(s,t,u)  u_{ \lambda_1,\, s_k} (p_1)=H^{H q\bar q g}_{\lambda_1\lambda_2\lambda_3} A^{H q\bar q g}_{\lambda_1\lambda_2\lambda_3}(s,t,u).
\eeq
Note that in the above equation we dropped color dependence for convenience.
The helicity prefactors $H^{H  q\bar qg}_{ \lambda_1\lambda_2\lambda_3}$ are defined by
\bea
H_{-+-}&=&H_{+-+}^*=- \frac{1}{\sqrt{2} } \frac{ \langle 13\rangle^2}{\langle 1 2 \rangle },\\
H_{-++}&=&H_{+--}^*=- \frac{1}{\sqrt{2} }\frac{ [ 32]^2}{[ 21  ] }.
\eea

The partial helicity amplitudes $A^{H  q\bar q g}_{ \lambda_1\lambda_2\lambda_3}(s,t,u)$ satisfy a perturbative expansion defined as in Eq.~\eqref{eq:pertexp}.
The tree-level amplitudes $A^{(0),\,H  q\bar q g}_{ \lambda_1\lambda_2\lambda_3}(s,t,u)$ are normalized to 1,
\beq
A^{(0),\, H  q\bar q g}_{ \lambda_1\lambda_2\lambda_3}(s,t,u)=1.
\eeq
Note that only two of the four non-vanishing helicity amplitudes are not identical.
\bea
A^{H  q\bar q g }_{-+-}&=&A^{H  q\bar q g }_{+-+},\nonumber\\
A^{H  q\bar q g }_{-++}&=&A^{H  q\bar q g }_{+--}.
\eea
Additionally, we find the symmetry
\bea
\label{eq:hqqbarhelamps}
A^{H  q\bar q g }_{-+-}(s,t,u)&=&
A^{H  q\bar q g }_{-++}(s,u,t).
\eea
The partial helicity amplitudes form a particularly convenient basis for the form factors because they satisfy the above relation.
We choose to present our final results in terms of the partial helicity amplitudes and present the invertible relation to the form factors in the conventional dimensional regularization representation here.
\beq
\left(\begin{array}{c}
A^{H  q\bar q g}_{-++}  \\ A^{H  q\bar q g}_{-+-}
\end{array}
\right)=
\left(
\begin{array}{cc}
 1 & 0 \\
 -\frac{t}{u} & \frac{t+u}{u} \\
\end{array}
\right)
\cdot
\left(\begin{array}{c}
A_1^{H  q\bar q g} \\ A_2^{H  q\bar q g}
\end{array}
\right).
\eeq

\section{Details on the Computation}
\label{sec:Computation}
\subsection{Integrand Construction}
The integrand is constructed starting from Feynman rules derived from the Lagrangian in Eq.~\eqref{eq:Lagrangian}.
We generate the Feynman diagrams for our scattering amplitudes with \textsc{qgraf}~\cite{qgraf}.
We then use our in-house {\textsc{C++}} packages to compute the amplitudes, including the contraction of the Lorentz and color indices.
Specifically, we project our amplitudes on the tensor structures we defined above in Eq.~\eqref{eq:tensdefggg} and Eq.~\eqref{eq:tensdefqqbarg}.
We perform color algebra using the public package \textsc{CIFAR}~\cite{Mistlberger:2025ksa}.
Next, we express all partial scattering amplitudes as linear combinations of scalar integrals and classify them into 219 integral families.

\subsection{Integral Reduction}
We use the public tool \textsc{CalcLoop}~\cite{calcloop} to identify external-momentum permutation symmetries of the Feynman integrals.
This reduces the number of integral families from 219 to 46 and the number of target integrals from 7 million to 0.25 million.
The amplitude is then written in terms of permutations of linear combinations of Feynman integrals within these 46 unique families.
The scalar integrals required to express our amplitude contain monomials of irreducible scalar products of up to rank 6.
For each distinct integral family, we perform integration-by-parts (IBP) reductions~\cite{Tkachov1981,Chetyrkin1981,Laporta:2000dsw} using the package \textsc{Blade}~\cite{Guan:2024byi}.
The IBP reduction is one of the main computational bottlenecks of this calculation.
A direct finite-field reconstruction from numerical IBP reductions~\cite{vonManteuffel:2014ixa,Peraro:2016wsq} would require solving very large linear systems at an impractically large number of phase-space points.
We therefore use the original IBP systems only to construct a block-triangular (BT) form~\cite{Liu:2018dmc,Guan:2019bcx}.
The resulting BT form is then used as a much faster evaluator for the large number of samples needed in the functional reconstruction~\cite{vonManteuffel:2014ixa,Peraro:2016wsq}.
The finite-field evaluation infrastructure is based on \textsc{FiniteFlow}~\cite{Peraro:2019svx}.
We also made practical modifications to \textsc{Blade} to reduce I/O overhead and to parallelize the computation over different prime fields.

We quote the cost for the most complicated topology, shown in Fig.~\ref{fig:3loop-topology}, to illustrate the computational bottleneck of the IBP reduction and the practical gain from using the BT form for finite-field reconstruction.\footnote{An example \textsc{Mathematica} setup for this benchmark is provided in the \textsc{Blade} repository at \href{gitee.com/multiloop-pku/blade}{\texttt{gitee.com/multiloop-pku/blade}}.}
This topology contains ten propagators, five irreducible scalar products, and 371 master integrals.
We consider the reduction of top-sector integrals at rank $r$; see Refs.~\cite{Smirnov:2008iw,Lee:2012cn,Maierhofer:2017gsa,Guan:2024byi} for the notation used in IBP reductions.
To reduce the size of IBP systems, we 
decompose the reduction into 29 spanning sectors~\cite{Larsen:2015ped,Guan:2024byi}.
The numerical IBP systems are generated with the improved seeding strategy of Refs.~\cite{Guan:2024byi,Driesse:2024xad,Bern:2024adl}, which assigns lower seed ranks to sectors with fewer propagators.
Combining the spanning-sector and improved seeding strategies keeps the memory usage of the numerical IBP reduction at the $\mathcal{O}(100) \, \mathrm{GB}$ level, making the reduction feasible on typical high-performance computing clusters..
The BT search also uses the mass dimensions of the Mandelstam variables to reduce the ansatz size and, consequently,  the number of direct numerical IBP probes.

\begin{table}
    \centering
    \small
    \setlength{\tabcolsep}{3pt}
    \renewcommand{\arraystretch}{1.5}
    \begin{tabularx}{\linewidth}{c*{6}{>{\centering\arraybackslash}X}}
    \toprule
       rank &
       \makecell{sum\\ $t_{\mathrm{IBP}}$} & \makecell{sum\\ $t_{\mathrm{BT}}$} & \makecell{max\\ $N_{\mathrm{search}}$} & \makecell{max\\ $N_{\mathrm{fit}}$} & \makecell{max\\ $N_{\mathrm{recon}}$} & \makecell{max\\ $N_{\mathrm{primes}}$}\\
    \midrule
       5 & 620 s & 6 s & 416 & 416 & 427134 & 7+1 \\
       6 & 3100 s & 31 s & 832 & 624 & 692301 & 8+1 \\
    \bottomrule
    \end{tabularx}
    \caption{Representative numerical cost and reconstruction parameters for the most complicated topology, shown in Fig.~\ref{fig:3loop-topology}.}
    \label{tab:f38_ibp_summary}
\end{table}

The numerical cost is summarized in Table~\ref{tab:f38_ibp_summary}.
The table separates the expensive direct IBP probes, used to construct the BT form, from the much cheaper BT evaluations, used to generate the large reconstruction data set.
Here $t_{\mathrm{IBP}}$ denotes the time spent solving the original numerical IBP systems at a single probe, while $t_{\mathrm{BT}}$ denotes the time spent evaluating the resulting BT forms.
Both times are summed over the 29 spanning sectors.
The remaining quantities are reported as the maximum values among the spanning sectors: $N_{\mathrm{search}}$ is the number of direct numerical IBP probes used in the initial BT search over the first finite field, $N_{\mathrm{fit}}$ is the number of direct numerical IBP probes needed per subsequent prime after the BT structure has been found, $N_{\mathrm{recon}}$ is the number of BT evaluations used for functional reconstruction per prime, and $N_{\mathrm{primes}}$ is the number of primes used for reconstruction plus validation.

For the rank-6 reduction, the most complicated spanning sector requires 832 direct numerical IBP probes in the first finite field to find the BT form, and 624 direct IBP probes in each subsequent finite field.
The reconstruction data are then generated from the BT form, requiring 692301 BT evaluations per finite field for the same spanning sector.
This separation is crucial in practice: across the spanning sectors, one direct numerical IBP probe costs 50--350 seconds, whereas one BT evaluation costs only 0.7--2 seconds with negligible memory overhead.
The rank-6 reconstruction is completed with 8 primes, together with one additional prime used for validation.
The rank-6 reduction required approximately $(3\times10^4)$ core-hours, while the complete reduction is estimated to have required several hundred thousand core-hours.
These results demonstrate that state-of-the-art techniques make such reductions feasible with moderate computational resources.

After the reductions are performed family by family, we further exploit external-momentum symmetry to reduce all permutations of master integrals in each family into a minimal set of master integrals, which contains 2185 elements.
At this stage, the amplitudes can be expressed in terms of this minimal basis by substituting in IBP tables and applying the permutations.
\textsc{Flint}~\cite{flint} and \textsc{Ratracer}~\cite{Magerya:2022hvj} are frequently used for intermediate symbolic manipulations.

\subsection{Master Integrals}
After the IBP reduction, we map the master integrals to a canonical basis.
The differential-equation setup follows the standard chain of Refs.~\cite{Kotikov:1991pm,Kotikov:1991hm,Gehrmann:1999as,Henn:2013pwa}.
A representative non-planar topology in this setup is illustrated in Fig.~\ref{fig:3loop-topology}.
For one-mass three-point kinematics at three loops, several integral subsets were computed previously in Refs.~\cite{Gehrmann:2024tds,DiVita:2014pza,Canko:2021xmn,Henn:2023vbd,Syrrakos:2023mor}, with recent function-space organization for one-leg-off-shell amplitudes given in Ref.~\cite{Gehrmann:2024tds}.
Additional integrals were computed by us for the purpose of Ref.~\cite{Guan:2025awp} and find their use in the results of this article.
\makeatletter
\pgfdeclaredecoration{qcdgluon}{coil}
{
  \state{coil}[switch if less than=%
    0.5\pgfdecorationsegmentlength+%
    \pgfdecorationsegmentaspect\pgfdecorationsegmentamplitude+%
    \pgfdecorationsegmentaspect\pgfdecorationsegmentamplitude to last,
               width=+\pgfdecorationsegmentlength]
  {
    \pgfpathcurveto
    {\pgfpoint@onqcdgluon{0    }{ 0.555}{1}}
    {\pgfpoint@onqcdgluon{0.445}{ 1    }{2}}
    {\pgfpoint@onqcdgluon{1    }{ 1    }{3}}
    \pgfpathcurveto
    {\pgfpoint@onqcdgluon{1.555}{ 1    }{4}}
    {\pgfpoint@onqcdgluon{2    }{ 0.555}{5}}
    {\pgfpoint@onqcdgluon{2    }{ 0    }{6}}
    \pgfpathcurveto
    {\pgfpoint@onqcdgluon{2    }{-0.555}{7}}
    {\pgfpoint@onqcdgluon{1.555}{-1    }{8}}
    {\pgfpoint@onqcdgluon{1    }{-1    }{9}}
    \pgfpathcurveto
    {\pgfpoint@onqcdgluon{0.445}{-1    }{10}}
    {\pgfpoint@onqcdgluon{0    }{-0.555}{11}}
    {\pgfpoint@onqcdgluon{0    }{ 0    }{12}}
  }
  \state{last}[next state=final]
  {
    \pgfpathcurveto
    {\pgfpoint@onqcdgluon{0    }{ 0.555}{1}}
    {\pgfpoint@onqcdgluon{0.445}{ 1    }{2}}
    {\pgfpoint@onqcdgluon{1    }{ 1    }{3}}
    \pgfpathcurveto
    {\pgfpoint@onqcdgluon{1.555}{ 1    }{4}}
    {\pgfpoint@onqcdgluon{2    }{ 0.555}{5}}
    {\pgfpoint@onqcdgluon{2    }{ 0    }{6}}
  }
  \state{final}{}
}
\def\pgfpoint@onqcdgluon#1#2#3{%
  \pgf@x=#1\pgfdecorationsegmentamplitude%
  \pgf@x=\pgfdecorationsegmentaspect\pgf@x%
  \pgf@y=#2\pgfdecorationsegmentamplitude%
  \pgf@xa=0.083333333333\pgfdecorationsegmentlength%
  \advance\pgf@x by#3\pgf@xa%
}
\makeatother
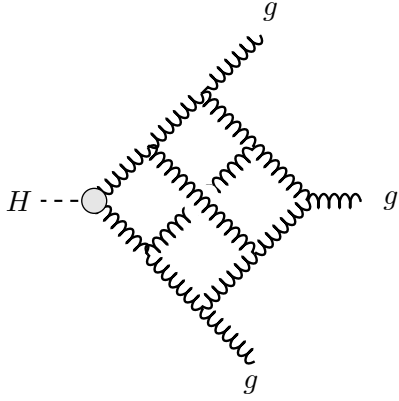
\begin{figure}[t]
    \centering
    \begin{tikzpicture}[scale=1.0,rotate=45]
        \tikzset{
            gline/.style={draw=black,line width=0.9pt,decorate,decoration={coil,aspect=0.35,segment length=1.4mm,amplitude=0.9mm}},
            hblob/.style={circle,fill=gray!20,draw=black,line width=0.5pt,minimum size=3.3mm,inner sep=0pt}
        }
        \coordinate (e1) at (-1.5,1.5);
        \coordinate (e2) at (2.1,1.0);
        \coordinate (e3) at (1.5,-1.5);
        \coordinate (e4) at (-1.0,-2.0);

        \coordinate (v1) at (-1,1);
        \coordinate (v2) at (0,1);
        \coordinate (v3) at (1,1);
        \coordinate (v4) at (1,0);
        \coordinate (v5) at (1,-1);
        \coordinate (v6) at (0,-1);
        \coordinate (v7) at (-1,-1);
        \coordinate (v8) at (-1,0);
        
        \draw[dashed,thick] (v1) -- (e1);
        \draw[gline] (e4) -- (v7);

        \fill[white] (v1) -- (v2) -- (v3) -- (v4) -- (v5) -- (v6) -- (v7) -- (v8) -- cycle;

        \draw[gline] (v3) -- (v4);
        \draw[gline] (v5) -- (v6);
        \draw[gline] (v6) -- (v7);
        \draw[gline] (v7) -- (v8);

        \draw[gline] (e2) -- (v3);
        \draw[gline] (v3) -- (v2);
        \draw[gline] (v2) -- (v1);
        \draw[gline] (v1) -- (v8);
        \draw[gline] (v8) -- (v4);
        \draw[gline] (v4) -- (v5);
        \draw[gline] (v5) -- (e3);

        \draw[draw=white,line width=2.4mm,line cap=round]
          ($(v8)!0.43!(v4)$) -- ($(v8)!0.57!(v4)$);

        \draw[gline] (v2) -- (v6);
        \node[hblob] at (v1) {};

        \node[fill=white,inner sep=1pt] at (2.42,1.12) {$g$};
        \node[fill=white,inner sep=1pt] at (-1.25,-2.18) {$g$};
        \node[fill=white,inner sep=1pt,right=7pt] at (e3) {$g$};
        \node[fill=white,inner sep=1pt,left=2pt] at (e1) {$H$};
    \end{tikzpicture}
    \caption{Representative non-planar three-loop topology, corresponding to the most involved integral family in Table~1 of Ref.~\cite{Guan:2025awp}.}
    \label{fig:3loop-topology}
\end{figure}

To obtain a canonical form we use a hybrid approach combining the Lee alogorithm \cite{Lee:2014ioa} with ideas introduced in \emph{CANONICA} \cite{Meyer:2017joq}. To this end the Lee algorithm is run on an univariate slice $(t,u)=(t_0+a \lambda,u_0+b\lambda)$ of the differential equation $A$. After this stage we can reconstruct the canonical form \cite{Henn:2013pwa} 
\begin{equation}
   \mathrm{d}\vec{\mathcal{I}}=C \vec{\mathcal{I}}\,,\qquad
    C= \epsilon \sum_{i=1}^{20} C_i\,\mathrm{d}\log(w_i)\,.\label{eq:diffeq}
\end{equation}
The alphabet is given by the 20 letters first identified in Ref.~\cite{Gehrmann:2024tds}
\begin{equation}
    w_i \in
    \left\{
      t,\ 1-t,\ t-(1-u)^2,\ 1-t-t u,\ \frac{r+t}{r-t}
    \right\}
    \ \cup\ \text{permutations in }(s,t,u)\,.
\end{equation}
The square-root $r$ is related to the  the non-planar integrals and explicitly given by
\begin{equation}
\label{eq:rootdef}
    r= \sqrt{-s\,t\,u}\,.
\end{equation}
From the evaluation on the slice we can guess the denominators as well as the numerator degree of $T$. Similar approaches are routinely applied for reconstructing amplitude coefficients (see e.g. Ref.~\cite{Abreu:2018zmy}). This leads to an small ansatz for $T$, the unknowns of which are determined by solving \cite{Meyer:2017joq}
\begin{equation}
    T C=AT-\mathrm{d}T\,.
\end{equation}
Compared to the Ref.~\cite{Meyer:2017joq} we have the additional benefit of knowing $C$ as well as having explicit degree bounds for the rational functions in $T$.
Additional details on the master integrals and their computation will be provided in a dedicated future publication.
We determine the boundary conditions to fully specify our solution to the differential equations exploiting regularity conditions and elementary integrals, see for example Refs.~\cite{Henn:2020lye,Dulat:2014mda,Henn:2013woa}.
We then solve the system by integrating it order-by-order in $\epsilon$ in terms of Chen iterated integrals~\cite{Chen:1977oja} in the variable $t$ and harmonic polylogarithms of $u$.
Permutations of external momenta act as a non-trivial rotation on our system of master integrals.
We exploit this as a cross-check on our system and compare the explicit change of variables induced by a permutation on our integrals to the linear combination generated by a permutation and find consistency.

The space of functions necessary to express our scattering amplitudes at three loops is significantly larger than the one required for lower orders, see Ref.~\cite{Gehrmann:2024tds} for a detailed discussion.
In particular, only the two letters $\{t,1-t\}$ and their permutations were required at lower loops~\cite{Gehrmann:2000zt,Dixon:2009uk,Badger:2006us,Badger:2009hw,Gehrmann:2023etk,Gehrmann:2011aa}.
All letters in our alphabet appear in functions contributing to our scattering amplitudes.
However, the alphabet can naturally be decomposed into two parts.

First, the part without any square root letters contains letters that are maximally quadratic in each variable.
We can therefore easily choose to factor the quadratic polynomials in one variable and rewrite our Chen iterated integrals in terms of more widely used generalized polylogarithms (GPLs)~\cite{Goncharov1998}, which can easily be manipulated~\cite{Duhr:2011zq,Duhr:2012fh} and numerically evaluated~\cite{zhenjiempl,Bauer2000,Vollinga:2004sn}.
\begin{equation}
    G(a_1,\ldots,a_n;t),\qquad n=1,\ldots,6,
\end{equation}
with GPL indices drawn from
\begin{align}
    a_i \in &\left\{\vphantom{\frac{1\pm\sqrt{1-4u}}{2}}
    0\,,
    1\,,
    -u\,,
    1-u\,,
    u(1-u)\,,
    (1-u)^2\,,\right.\nonumber\\
&  \left. \qquad 1\pm\sqrt{u},\,
    -u\pm\sqrt{u},\,
    \frac{1\pm\sqrt{1-4u}}{2},\,
   \frac{1\pm\sqrt{1-4u}-2 u}{2}
    \right\}.
\label{eq:ai-alphabet}
\end{align}

The second subset of our functions involves the letters with the square root $r$.
We note that for iterated integrals involving root letters, only $t$, $1-t-u$, and the two independent square-root letters appear.
We find that in the scattering amplitudes functions depending on $r$ have a maximal transcendental weight of five.
Here, we refer to transcendental weight as the number of iterated integrations necessary to compute a function from our differential equations, Eq.~\eqref{eq:diffeq}.
We rationalize the square-root by introducing the change of variables
\begin{equation}
    t=-\frac{(1-u)z^2}{u-z^2}\,.
\end{equation}
This covers the real interval $t\in[0,1]$ twice and we choose the sheet $\operatorname{Re}(z)=0,\,\operatorname{Im}(z)>0$, such that
\begin{equation}
\label{eq:zdef}
    z=\imath \sqrt{\frac{tu}{s}}=\imath\frac{\sqrt{t}\sqrt{u}}{\sqrt{1-t-u}}\,,\quad  r= \frac{u(1-u)z}{u-z^2}\,.
\end{equation}
In the rational parametrization the form factor can be written in terms of generalized polylogarithms
\begin{equation}
    G(b_1,\ldots,b_n;z),\qquad n=1,\ldots,5,
\end{equation}
with GPL indices
\begin{equation}
    b_i \in \{0,\,\pm1,\,\pm\sqrt{u},\,\pm u\}.
\label{eq:bi-alphabet}
\end{equation}
The translation strategy is: rationalize, extract all logarithms in $t$, and rewrite the remaining kernels in $z$. After partial-fraction decomposition, poles map directly to GPL indices and the result is integrated recursively in weight.
As an explicit example of the reduction to GPLs, we obtain
\begin{align}
    \mathcal{I}\!\left[\frac{r+u}{r-u},t\right]\!(t)
    ={}&\int_0^t\df t^\prime \frac{\df }{\df t^\prime} \log\left(\frac{r+u}{r-u} \right) \int_1^{t^\prime} \df t^{\prime \prime} \frac{\df \log (t^{\prime \prime} )}{\df t^{\prime \prime} }.
    \nonumber\\
    ={}&    2\,G(0,-1;z)-2\,G(0,1;z)-2\,G(0,-u;z)+2\,G(0,u;z)\nonumber \\
& -G(-\sqrt{u},-1;z)+G(-\sqrt{u},1;z)+G(-\sqrt{u},-u;z)-G(-\sqrt{u},u;z) \nonumber \\
& -G(\sqrt{u},-1;z)+G(\sqrt{u},1;z)+G(\sqrt{u},-u;z)-G(\sqrt{u},u;z) \nonumber \\
& +\bigl[-G(-1;z)+G(1;z)+G(-u;z)-G(u;z)\bigr]\log t\,.
\end{align}
All Chen-to-GPL replacements were checked by matching series expansions around $t=0$. Notice that while obscured in this representation, the function is analytic at $u=0$ by construction. Given the expression in terms of GPLs the form factors can be evaluated efficiently e.g. using~\cite{zhenjiempl,Bauer2000,Vollinga:2004sn}. When evaluating in scattering kinematics, e.g. $s>0,t,u<0$, the correct analytic continuation is ensured by adding a positive small imaginary part to $t$ and $u$ following the Feynman prescription.

It is instructive to study the space of functions in our scattering amplitude.
Specifically, we count the number of independent functions of pure transcendental weight with only integer coefficients $ \mathcal{F}_i(s,t,u)$ which are required to express our partial amplitudes, such that each amplitude can be written as
\beq
\mathcal{A}^{(n)}=\sum_i P_i^{(n)}(s,t,u) \mathcal{F}_i(s,t,u),
\eeq
where the $P_i^{(n)}(s,t,u)$ are ratios of polynomials or algebraic functions of our variables.
\begin{table}[h!]
\centering
\small
\setlength{\tabcolsep}{3pt}
\renewcommand{\arraystretch}{1.5}
\begin{tabular}{lcccccc}
\toprule
weight & 1 & 2 & 3 & 4 & 5 & 6\\
\midrule
\# of Functions & 4 & 14 & 45 & 129 & 97 & 5\\
\bottomrule
\end{tabular}
\caption{Number of pure transcendental functions at a given weight required to express all our partial helicity amplitudes.}
\label{tab:funcnums}
\end{table}
The number of these functions is reported in Table~\ref{tab:funcnums}.
For example, the weight one functions are simply the logarithms of $\{s,t,u,-q^2 / \mu^2\}$.

\subsection{Analytic Continuation}

Our scattering amplitudes can find a range of physical and formal applications and, depending on the pursued purpose, they will have to be evaluated in different external kinematic regions.
We performed our calculations in an unphysical region in which we treated all invariants to be negative -- the pseudo-Euclidean region.
The advantage of choosing this region is that all our functions are manifestly real-valued in the triangle spanned by
\beq
t,u \in [0,1],\hspace{1cm} s=1-t-u<1,\hspace{1cm}q^2< 0.
\eeq
This region allows us to study the function space of the scattering amplitudes and draw comparisons to computations in $\mathcal{N}=4$ super Yang-Mills theory.
However, physical scattering experiments are performed in different regions.
Higgs boson production at the Large Hadron Collider, Higgs exchange in Deep Inelastic Scattering and the decay of a Higgs boson after production (for example at a future muon collider) come to mind.
Analytic continuation of our functions can be performed by crossing their physical branch cuts, which are located solely where a Mandelstam invariant (including $q^2$) is 0.
The three kinematic regimes of a Higgs boson $H$ interacting with three partons $q_i$ are characterized as follows.
\subsubsection*{Decay: $H\to q_1+q_2+q_3$}
\beq
q^2>0, \hspace{1cm}s,t,u>0.
\eeq
Analytic continuation to this region is relatively simple as only $q^2$ changes sign in comparison to the Euclidean region.
This can be achieved analytically by replacing
\beq
\log(-q^2)=\log(q^2) - \imath\pi.
\eeq
\subsubsection*{Production: $q_1+q_2 \to H +q_3$}
The production kinematics of a Higgs boson in the annihilation of two partons is specified by the following region.
\beq
q^2>0,\hspace{1cm}
s>0,\hspace{1cm}
t<0,\hspace{1cm}
u<0.
\eeq
First, analytic continuation is dictated by the Feynman prescription assigned to each propagator.
This leads to the fact that each Mandelstam invariant should be thought of as carrying an infinitesimal imaginary part.
\beq
(p_i+p_j)^2\to (p_i+p_j)^2 +\imath \delta,\hspace{1cm} q^2\to q^2+\imath \delta.
\eeq
This on its own fixes the analytic continuation of our functions, and indeed simply evaluating the GPLs defined in the Euclidean region with very small explicit imaginary parts with publicly available tools leads to consistent results.
\subsubsection*{Deep Inelastic Scattering: $q_1+H\to q_2 +q_3$}
Lastly, one could consider the exchange of a virtual Higgs boson in the collisions of a lepton and a hadronic target as in Deep Inelastic Scattering.
While this scattering process is very hard to observe we nevertheless note its kinematic regime here.
\beq
q^2<0,\hspace{1cm}
s>0,\hspace{1cm}
t<0,\hspace{1cm}
u>0.
\eeq
Analytic continuation for numerical evaluation is again most easily performed by equipping invariants with an infinitesimal (small) imaginary part.

\subsubsection*{The Square Root}
Particularly curious is our square root $r$, defined in Eq.~\eqref{eq:rootdef}.
If its argument is negative, selecting a branch to take the square root numerically should be specified by the Feynman prescription as well.
In particular, in DIS kinematics the argument of our root changes sign and a prescription for analytic continuation seems necessary.
However, we verify that our amplitude enjoys a Galois symmetry that allows us to choose the branch of the square root arbitrarily, and both possible branches lead to the same numerical result as long as the root is treated consistently through the analytic functions and their algebraic prefactors.
Specifically, when one of the three variables $\{s,t,u\}$ is negative the variable $z$ introduced in Eq.~\eqref{eq:zdef} turns real.
In DIS kinematics this is the variable $t$.
\beq
z=\imath \sqrt{\frac{ut}{s}}=\imath\sqrt{\frac{-(p_2+p_3)^2-\imath\delta}{-q^2-\imath \delta} \frac{u}{s}}=\imath \sqrt{(t-\imath \delta) \frac{u}{s}}=\sqrt{\frac{(-t) u }{s}}.
\eeq
However, choosing $z\to -\sqrt{\frac{(-t) u }{s}}$ would yield identical results, as long as done consistently. 
In general, our results are invariant under $z\to - z$.

\section{Infrared Structure \label{sec:IR}}

Infrared and ultraviolet singularities of scattering amplitudes manifest as poles in the dimensional regulator $\epsilon$.
The poles of an amplitude at order $n$ in perturbation theory are described by the following universal formula in terms of its lower order amplitudes such that we can define a finite remainder $\mathcal{A}_X^F$.

\bea
\label{eq:AFdef}
\mathcal{A}_X^F&&={\mathbf Z}(\alphas(\mu^2),\{p_i\})\mathcal{A}_X(\alphas,\,C^0)\nonumber\\
&&={\mathbf Z}(\alphas(\mu^2),\{p_i\})\mathcal{A}_X({\mathbf Z}_{\alphas}N_\epsilon\alphas^R,\,{\mathbf Z}_{C^0}C^{0,R}).
\eea

Here, $\alphas^R$ and $C^{0,R}$ are the renormalized strong coupling and Wilson coefficient, the operator ${\mathbf Z}_{\alphas}$ and ${\mathbf Z}_{C^0}$ implement the $\overline{\text{MS}}$ ultraviolet renormalization of the strong coupling constant and the effective operator of Eq.~\eqref{eq:Lagrangian} using the $\beta $-function of QCD~\cite{Baikov:2016tgj,Herzog:2017ohr,Czakon:2004bu,vanRitbergen:1997va,Larin:1993tp,Tarasov:1980au,Kniehl:2006bg} and the IR singularities are subtracted by the universal factor ${\mathbf Z}(\alphas(\mu^2),\{p_i\})$~\cite{Almelid:2015jia,Aybat:2006mz,Aybat:2006wq,Catani:1998bh,Dixon:2008gr,Korchemsky:1987wg,Sterman:2002qn,Becher:2019avh}.
\beq
\label{eq:Zexp}
{\mathbf Z}(\alphas(\mu^2),\{p_i\},\epsilon) = \mathcal{P} e^{\-\frac{1}{4}\int_{0}^{\mu^2} \frac{d\mu^{\prime 2}}{\mu^{\prime 2}} {\mathbf \Gamma(\alphas(\mu^{\prime 2}),\{p_i \}, \epsilon)} }\,,
\eeq
with
\bea
\label{eq:GammaSoftDef}
 {\mathbf \Gamma(\alphas(\mu^2),\{p_i \}, \epsilon)}&& = \sum_{i\neq j} {\mathbf T}_i^a {\mathbf T}_j^a \Gamma_{\text{cusp}}(\alphas(\mu^2)) \log \frac{-s_{ij}}{\mu^2} \nonumber\\
 &&+ \frac{1}{2}\sum_i \mathbf{\mathds{1}} \gamma^{R_i}_c+\bold{\Delta}(\alphas(\mu^2),\{p_i \}).
\eea
Above, the symbol $\Gamma_\text{cusp}$ refers to the cusp anomalous dimension~\cite{Korchemsky:1987wg}, which is currently known exactly through four-loop order~\cite{Henn:2019swt,vonManteuffel:2020vjv}, and approximately at five loops \cite{Herzog:2018kwj}.
Furthermore, and $\gamma_c^R$ is the collinear anomalous dimension, obtained through four-loop accuracy in Refs.~\cite{Agarwal:2021zft,vonManteuffel:2020vjv}. The above formula was derived and calculated to three-loop order in Ref.~\cite{Almelid:2015jia} and verified in $\mathcal{N}=4$ super Yang-Mills theory~\cite{Henn:2016jdu} and QCD~\cite{Caola:2022dfa,Caola:2021izf,Caola:2021rqz}.
Furthermore, in Ref.~\cite{Becher:2019avh}, its general structure was determined to four-loop order.
The term $\bold{\Delta}(\alphas(\mu^2),\{p_i \})$ is known as the correction of the dipole formula and starts at three-loop order.
We refer, for example, to Sec.~5 of Ref.~\cite{Herzog:2023sgb} for further details.

We verified that the poles of our scattering amplitude through three-loop order match the analytic functions predicted by Eq.~\eqref{eq:AFdef} identically.
This is one of the most stringent checks on our results and furthermore allows us to extract the finite remainder of our partial scattering amplitudes $A_i^{F}$.

\section{Results \label{sec:Results}}
In this section, we discuss briefly the numerical impact of the finite part of our scattering amplitudes.
We consider the interference with the tree-level amplitude,
\begin{equation}
\mathcal{M}^{(n)}_X=2 \text{Re} \sum_{\text{helicities, color}}\mathcal{A}^{(n)}_X\mathcal{A}^{\dagger\,(0)}_X.
\end{equation}
We find specifically,
\bea
\mathcal{M}^{(n)}_{H ggg}&=& g_{\mathrm{s}}^2 (C^{0})^2 \frac{4C_A^2 C_F q^2}{s t u}  \text{Re} \left[ A^{F\,(n)\,H ggg}_{+++} + s^4  A^{F\,(n)\,H ggg}_{++-}+ u^4  A^{F\,(n)\,H ggg}_{+-+} + t^4  A^{F\,(n)\,H ggg}_{-++}\right], \nonumber\\
\mathcal{M}^{(n)}_{H q\bar q g}&=& g_{\mathrm{s}}^2 (C^{0})^2  2C_A C_F q^2 \text{Re} \left[\frac{u^2}{s} A^{F\,(n)\,H q\bar qg}_{-+-} + \frac{t^2}{s}  A^{F\,(n)\,H q\bar qg}_{-++} \right].
\eea

\begin{table}[h!]
\centering
\small
\setlength{\tabcolsep}{3pt}
\renewcommand{\arraystretch}{1.5}
\begin{tabular}{lcccc}
\toprule
Channel & $(s,t,u)$ & $\mathcal{M}^{(1)}_X/\mathcal{M}^{(0)}_X $ & $\mathcal{M}^{(2)}_X/\mathcal{M}^{(0)}_X$ & $\mathcal{M}^{(3)}_X/\mathcal{M}^{(0)}_X$ \\
\midrule
$H\!\to\! ggg$ & $\left(\frac{1}{3},\frac{1}{3},\frac{1}{3}\right)$ & -2.33105  & $ - 16.2085 $ & $-237.487$ \\
$H\!\to\! q\bar q g$ & $\left(\frac{1}{3},\frac{1}{3},\frac{1}{3}\right)$ & $1.43795$  & $2.43399$ & $33.4291$ \\
\midrule
$gg \to Hg$ & $\left(2,-\frac{1}{3},-\frac{2}{3}\right)$ &
$14.3033$ &
$148.012$ &
$1399.67$ \\
$ q\bar q \to H g$ &
$\left(2,-\frac{1}{3},-\frac{2}{3}\right)$ &
$11.6416$ &
$133.567$ &
$1563.97$ \\
\bottomrule
\end{tabular}
    \caption{Reference values for the interference terms crossed to various channels. We have set $q^2=-\mu^2=1$ and $N_c=3$, $N_f=5$. The numerical accuracy was truncated at the last digit.
}
\label{tab:NumericValues}
\end{table}
Numerical points that can serve as validation of our results are shown in Table~\ref{tab:NumericValues}.
We choose to evaluate our scattering amplitudes in the Euclidean region and the scattering region, setting $q^2=-\mu^2=1$.
We normalize our interfered scattering amplitudes by their tree-level value $\mathcal{M}^{(0)}_X$.

Next, we are interested in the perturbative behavior of our scattering amplitude.
To this end, we define the cumulant of the perturbative orders of the scattering amplitude interfered with the tree-level scattering amplitude.
\begin{equation}
    \mathcal{R}^{(n)}=\sum_{k=0}^n a_{\mathrm{s}}^k\operatorname{Re}\left(\frac{\mathcal{M}^{(k)}
    }{\mathcal{M}^{(0)}}\right).
\end{equation}

\begin{figure}[t]
    \centering
        \includegraphics[width=0.49\textwidth]{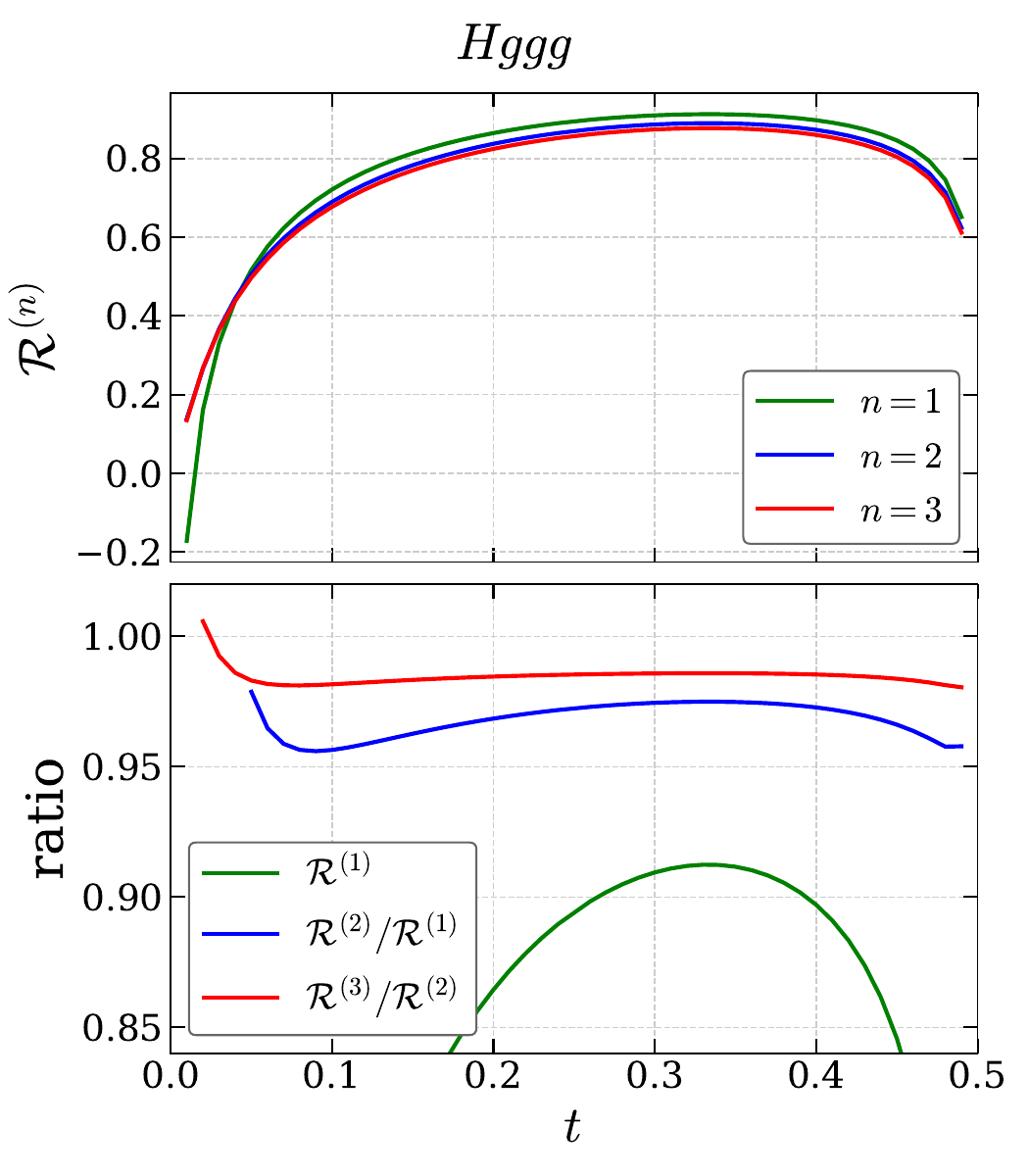}\includegraphics[width=0.49\textwidth]{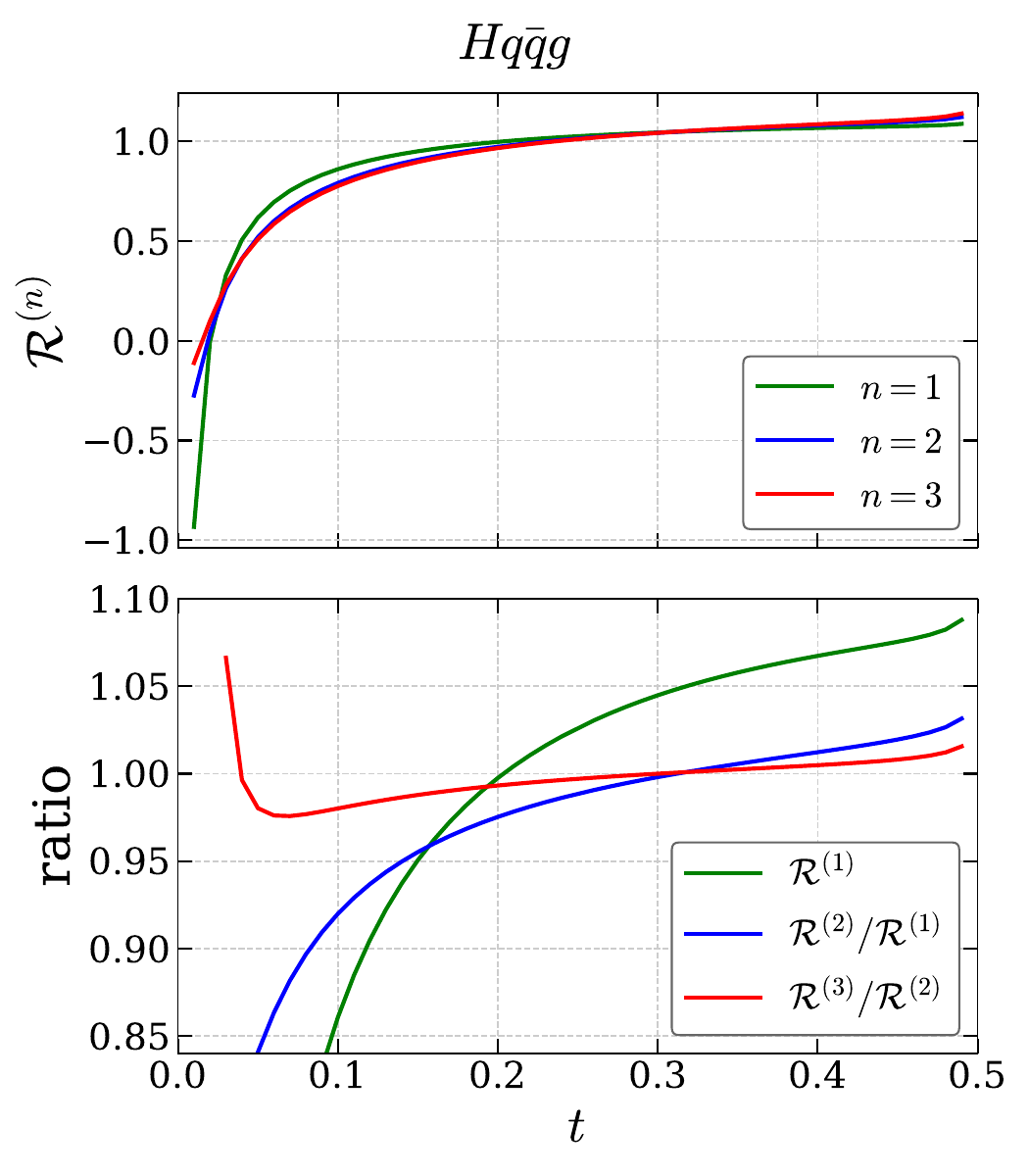}
    \caption{
    The ratio $\mathcal{R}^{(n)}$ on a euclidean kinematic slice $t=u$, $q^2=-\mu^2$ and $N_c=3$, $N_f=5$.
    The left panels show the ratio for the $H ggg$ channel and the right panels the ratio for the $H q\bar q g$ channel.
    The green, blue and red lines represent the ratio truncated at NLO, NNLO and N$^3$LO, respectively.
    The upper panels show the nominal value and the lower panels the ratio of two consecutive orders.
    }
    \label{fig:Rn}
\end{figure}
The cumulant $\mathcal{R}^{(n)}$ on the line $t=u$ in the interval $t\in [0,0.5]$ in the Euclidean region is shown in Fig.~\ref{fig:Rn}, with $q^2=
-\mu^2$.
We choose $\alphas(\mu^2)=0.118$, which corresponds to the scale of the $Z$ boson mass and is a representative choice for high-energy phenomenology processes.
Lines in green, blue and red correspond to truncating the perturbative expansion at NLO, NNLO and N$^3$LO, respectively.
While the quantity we are investigating is not a physical object, it still gives an indication towards the overall perturbative behavior of the scattering amplitude and its contribution to physical cross sections.
We observe in the lower panels of Fig.~\ref{fig:Rn} that our newly computed N$^3$LO contributions are overall smaller than previous orders, in accordance with our expectation that the series should behave perturbatively at this order.

\begin{figure}[h!]
    \centering
        \includegraphics[width=0.49\textwidth]{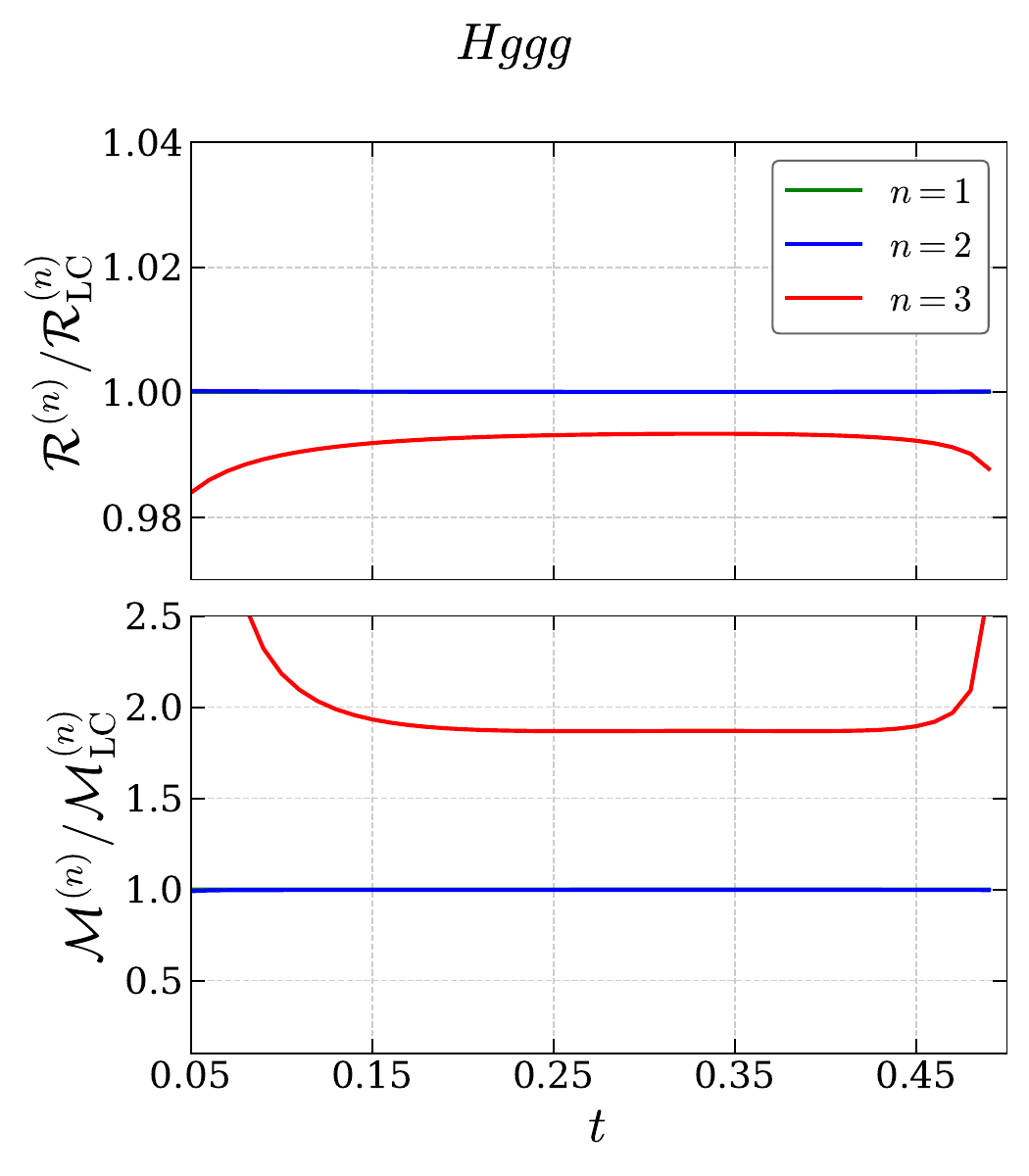}\includegraphics[width=0.49\textwidth]{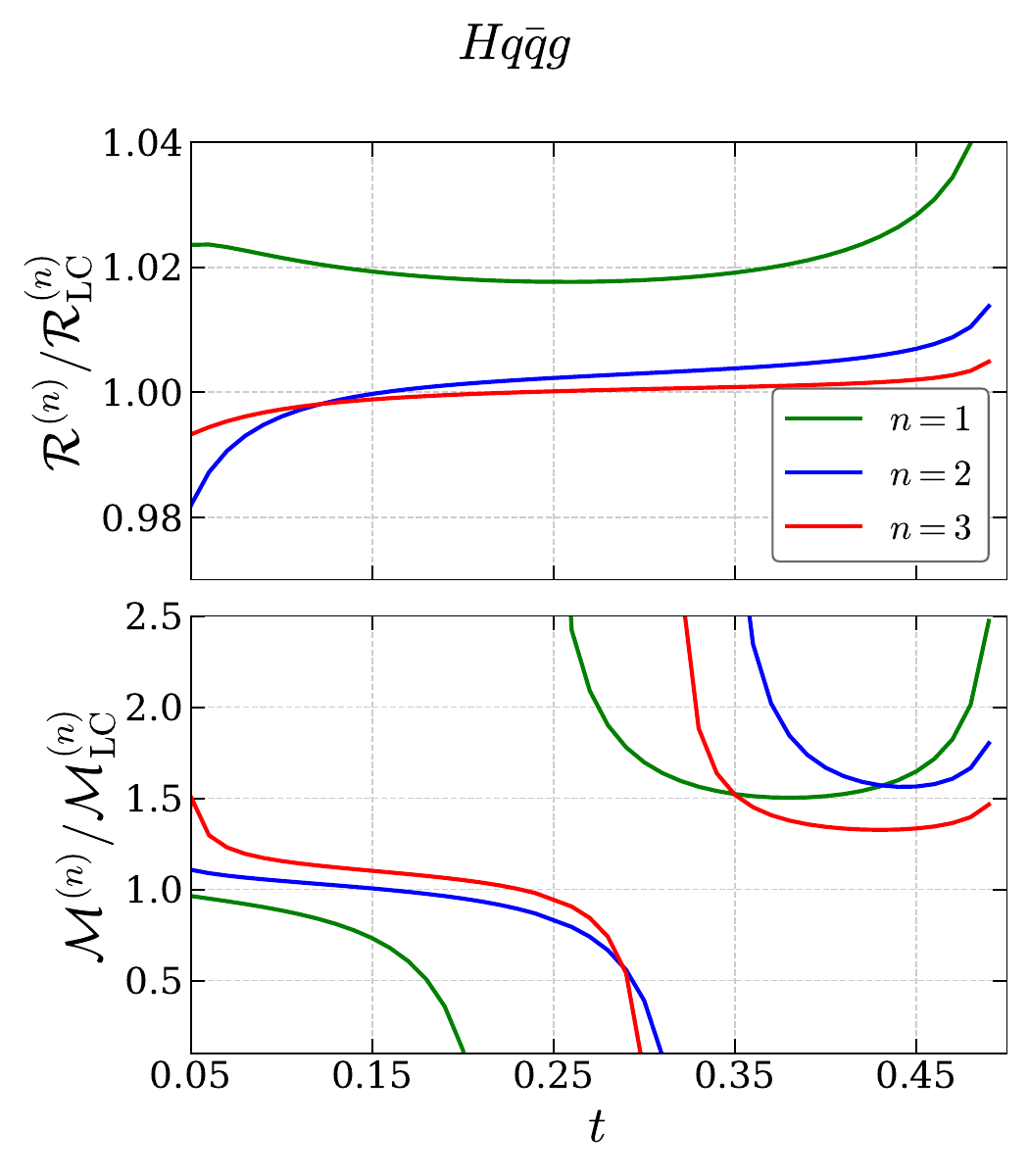}
    \caption{
     The upper panels show the ratio  $\mathcal{R}^{(n)}/\mathcal{R}^{(n)}_{\text{LC}}$ on a euclidean kinematic slice $t=u$, $q^2=-\mu^2$ and $N_c=3$, $N_f=5$.
     The lower panels show the ratio  $\mathcal{M}^{(n)}/\mathcal{M}^{(n)}_{\text{LC}}$ on the same kinematic slice.
    The left panel shows the ratio for the $H ggg$ channel and the right panel the ratio for the $H q\bar q g$ channel.
    The green, blue and red lines represent the choices $n=1$, $n=2$ and $n=3$, respectively.
    For the left plot, the blue and green line are congruent.
    }
    \label{fig:lc}
\end{figure}
The generalized leading-color approximation, which retains only the leading power in the number of colors $N_c$ or fermion flavor $N_f$ at a given perturbative order, is compared to the full computation achieved here in Fig.~\ref{fig:lc}.
Specifically, we include in the cumulant $\mathcal{R}_{\text{LC}}^{(n)}$ full-color dependence in all but the last order to study the impact of the new terms specifically computed in this article.
The upper panels show that the inclusion of subleading-color terms in our amplitudes changes the cumulant for both processes at the sub-percent level. However, this is rather a statement about the nominal contribution of three-loop contributions to the perturbative series truncated at N$^3$LO.
The lower panels of the plots of Fig.~\ref{fig:lc} demonstrate clearly that leading- and subleading-color approximations yield significant differences for the N$^3$LO coefficient of our ratio.
At NLO there are no subleading-color contributions for the $H ggg$ channel and at NNLO they are small enough to make the line in the left-bottom plot of Fig.~\ref{fig:lc} indistinguishable from unity.
At N$^3$LO the nominal correction due to subleading-color effects is of the order of the leading-color contribution.

\section{Validation \label{sec:Validation}}
In order to validate our results we have performed several checks.
\begin{enumerate}
\item
We verify that the poles of our scattering amplitudes correspond to the prediction discussed in Sec.~\ref{sec:IR}.
\item
We have performed an independent numerical check of our computation. Starting from the Feynman-diagram integrand, we evaluated the integrals numerically with AMFlow~\cite{Liu:2017jxz,Liu:2022chg} to obtain the value of the amplitude at the symmetric point $s=t=u=1/3$.
Consequently, we used the differential equation to perform numerical evaluations in other kinematic regions, thereby checking the analytic continuation provided through the GPL expression.
This provides a stringent check of the evaluation of the master integrals.
\item
We have studied the kinematic limit where two partons become collinear.
Generalized collinear factorization dictates that the scattering amplitudes have to factor into the two-parton form factor and universal splitting amplitudes, which were computed in Refs.~\cite{Gehrmann:2010ue,Lee:2022nhh,Baikov:2009bg,Gehrmann:2010tu,Gehrmann2005,Guan:2024hlf}.
The limit in which one gluon becomes soft is similarly described by factorization onto the three-loop soft current computed in Ref.~\cite{Herzog:2023sgb}. However, the soft limit is also obtained from the collinear limit and therefore does not provide an additional check.
\item
It was shown in Ref.~\cite{Brandhuber:2012vm} that the leading transcendental part of the scattering amplitude of a Higgs and three partons in QCD is identical to the $\operatorname{tr}(\phi^2)$ form factor in maximally supersymmetric Yang-Mills theory ($\mathcal{N}=4$ sYM).
This surprising result was recently confirmed to hold at three-loop order in the leading-color approximation in Ref.~\cite{Chen:2025utl}.
Here, we compare the leading transcendental contribution of all of our partial helicity amplitudes for a Higgs boson and three partons (both quarks and gluons) with full-color dependence to our recent results for the full-color form factor in $\mathcal{N}=4$ sYM theory.
To this end, we also choose the representation of our quarks to be the adjoint representation.
First, we find that all our partial helicity amplitudes are identical for this choice and that they indeed correspond to the form factor in $\mathcal{N}=4$ sYM theory, including the subleading-color contributions.
\item The generalized leading-color limit of our scattering amplitudes interfered with their tree-level counterpart was previously obtained in Ref.~\cite{Chen:2025utl}, and we find perfect agreement. Note that the results of Ref.~\cite{Chen:2025utl} were obtained by an overlapping subset of the authors and there is partial overlap in the computational ingredients.
\end{enumerate}

\section{Conclusions}
\label{sec:Conclusions}
In this article, we have presented three-loop QCD scattering amplitudes for a Higgs boson and three light partons in the heavy-top effective field theory.
Through analytic continuation, these amplitudes provide essential ingredients for physical predictions in high-energy particle physics phenomenology.
For example, our amplitudes are virtual corrections for the decay of a Higgs boson into three jets and for Higgs boson production at hadron colliders in association with a jet.

The results are provided in a compact analytical form using generalized polylogarithms.
This representation enables fast numerical evaluation sufficient for precision phenomenology.
We provide our results in terms of electronically readable files~\cite{ZenodoResults}.
Specifically, the files contain the bare partial helicity amplitudes of Eq.~\eqref{eq:hggghelamps} and Eq.~\eqref{eq:hqqbarhelamps} as well as their finite (IR-subtracted and UV-renormalized) counterparts as defined in Eq.~\eqref{eq:AFdef}.

We have performed a variety of structural, numerical and consistency checks of our results, summarized in Sec.~\ref{sec:Validation}.
We want to highlight the perhaps surprising observation that the leading transcendental part of all of our scattering amplitudes is identical among all amplitudes and corresponds to the $\operatorname{tr} (\phi^2)$ form factor computed in maximally supersymmetric Yang-Mills theory in Ref.~\cite{Guan:2025awp}.

The generalized leading-color limit of the scattering amplitudes computed here, interfered with their tree-level counterpart, was already computed in Ref.~\cite{Chen:2025utl}.
We find agreement with the result in the literature and study the impact of subleading-color contributions.
First, we observe that the space of functions, which was identified in Ref.~\cite{Gehrmann:2024tds}, has a much richer structure for subleading-color contributions due to a non-vanishing square root letter in the alphabet of generalized polylogarithms.
Second, we study the numerical impact of subleading-color contributions to the scattering amplitudes in the Euclidean scattering region.
We find that the numerical contribution of the subleading-color terms at three loops is comparable to the leading-color contribution.
This is in contrast to lower loop amplitudes.
However, we also observe that the impact of three-loop contributions is small assuming a value for the strong coupling constant typical for high-energy scattering processes.

Recently, the two-loop amplitudes for a Higgs boson and four partons have been computed in the heavy-top effective theory in the generalized leading-color approximation~\cite{Hartanto:2026xjz,DeLaurentis:2026brm}.
Together with a future full-color computation of these results, our results here yield the completion of the required scattering amplitudes to perform phenomenological computations of the production cross section of a Higgs boson and hadronic jets via the gluon-fusion mechanism at N$^3$LO in perturbative QCD.
Furthermore, our scattering amplitudes will enable high-precision predictions for Higgs boson phenomenology with jets in the final states at future muon colliders or electron-positron colliders.
Finally, our results will form a key ingredient towards the determination of the inclusive gluon fusion production cross section at N$^4$LO in perturbative QCD.

\section{Acknowledgments}
We thank Xiang Chen for his collaboration at the early stage of this project. We would also like to thank Lance Dixon for useful discussions.
MR, XG and BM are supported by the United States Department of Energy, Contract DE-AC02-76SF00515.

\addcontentsline{toc}{section}{References}
\bibliographystyle{jhep}
\bibliography{refs}

\end{document}